\documentclass[twocolumn,showpacs,superscriptaddress,preprintnumbers,amsmath,amssymb,prb]{revtex4}
\usepackage{graphicx}% Include figure files
\usepackage{dcolumn}% Align table columns on decimal point
\usepackage{bm}% bold math
\usepackage{amsfonts}% bold math
\usepackage[below]{placeins}
\begin{document}

%\preprint{APS/123-QED}

\title{Structural and magnetic properties of the new cobaltate series (BaSr)$_{4-x}$La$_{2x}$Co$_{4}$O$_{15}$}

\author{V. O. Garlea}
 \email{garleao@ornl.gov}
\affiliation{Quantum Condensed Matter Division, Oak Ridge National Laboratory, Oak Ridge, TN 37831, USA}
\author{R. Jin}
\affiliation{Department of Physics and Astronomy, Louisiana State University, Baton Rouge, LA 70803, USA}
\author{E. Garlea}
\affiliation{Development Division, Y12 National Security Complex, Oak Ridge, TN 37831, USA}
\author{G. Ehlers}
\affiliation{Quantum Condensed Matter Division, Oak Ridge National Laboratory, Oak Ridge, TN 37831, USA}
\author{E. Mamontov}
\affiliation{Chemical \& Engineering Materials Division, Oak Ridge National Laboratory, Oak Ridge, TN 37831, USA}
\author{D. B. Myers }
\affiliation{Department of Physics and Astronomy, Louisiana State University, Baton Rouge, LA 70803, USA}
\author{F. Xie}
\affiliation{School of Chemistry and Chemical Engineering, South
China University of Technology, Guangzhou 510006, China}
\author{R. Custelcean}
\affiliation{Chemical Sciences Division, Oak Ridge National Laboratory, Oak Ridge, TN 37831, USA}

\date{\today}% It is always \today, today,

\begin{abstract}
We report the structural and magnetic properties of a new class of cobaltates with the chemical formula (BaSr)$_{4-x}$La$_{2x}$Co$_{4}$O$_{15}$ ($x$ = 0, 0.5 and 1). These compounds crystallize in a hexagonal structure in which cobalt ions are distributed among two distinct crystallographic sites with different oxygen coordination. Three Co--O tetrahedra and one octahedron are linked by shared oxygen atoms to form Co$_4$O$_{15}$ clusters, which are packed together into a honeycomb-like network. Partial substitution of Sr and/or Ba atoms by La allows one to adjust the degree of Co valence mixing, but all compositions remain subject to a random distribution of charge. Magnetic susceptibility together with neutron scattering measurements reveal that all studied specimens are characterized by competing ferro- and antiferro-magnetic exchange interactions that give rise to a three dimensional Heisenberg spin-glass state. Neutron spectroscopy shows a clear trend of slowing down of spin-dynamics upon increasing La concentration, suggesting a reduction in charge randomness in the doped samples.
\end{abstract}


\maketitle

\section{Introduction}

Cobalt-based oxide compounds continue to attract a great deal of interest due to the richness of the magnetic and electronic properties given by strong correlation between spin, charge, and orbital degrees of freedom. A signature characteristic of cobalt oxides that distinguishes them from the other transition metal oxides is that Co ions are all susceptible to having multiple valence and/or spin states.~\cite{Yamada,Moritomo,Zaliznyak,Chichev,Senaris,Asai,Caciuffo}
The different electronic configurations can appear as a result of the interplay between the crystal field, caused by the local atomic environment, and the interatomic exchange interaction.
Perhaps the most extensively studied cobaltates in which the polyvalency and multi-spin states come to play to produce complex phase diagrams have been La$_{2-x}$Sr$_x$CoO$_4$,~\cite{Yamada,Moritomo,Zaliznyak,Chichev} and La$_{1-x}$Sr$_x$CoO$_3$,~\cite{Senaris,Asai,Caciuffo,Phelan}. These compounds possess perovskite-based structures with cobalt ions residing in an octahedral environment. In recent years, research has been expanded to include cobaltates containing Co$^{4+}$ or Co$^{3+}$ in tetrahedral sites, such as the Ba$_{2}$CoO$_{4}$,~\cite{Jin,Boulahya}, the oxocobaltates Na$_{6}$Co$_{2}$O$_{6}$, Na$_{5}$CoO$_{4}$,~\cite{Sofin} and the geometrically frustrated systems $Ln$BaCo$_{4}$O$_{7}$ (where $Ln$= lanthanides).~\cite{Valldor,Maignan,Chapon,Soda,Caignaert} As a result, new and interesting physics has continued to emerge as each of these systems display various exotic magnetic phases. Compounds that exhibit Co$^{3+}$ in both octahedral and tetrahedral lattice sites are less often found, and typically rely on an accurate control of the oxygen content (e.g. the oxygen-deficient perovskites Sr$_3$YCo$_4$O$_{10+\delta}$~\cite{Withers,Istomin,Khalyavin}).

\begin{figure}[btp]
\includegraphics[width=3.4in]{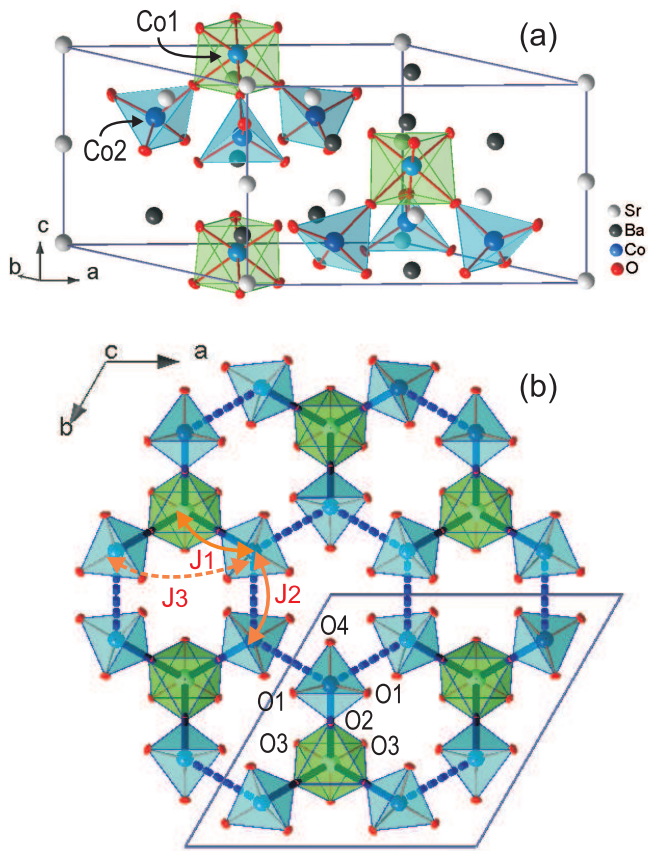}
\caption{\label{struct}(Color online) (a) Perspective view of the crystal structure of the (BaSr)$_{4}$Co$_{4}$O$_{15}$ compound.(b) Honeycomb-like layers of Co atoms projected along the crystallographic $c$ direction.}
\end{figure}

In this article, we consider yet another class of cobaltates, with the chemical formula (BaSr)$_{4-x}$La$_{2x}$Co$_{4}$O$_{15}$ ($x$ = 0, 0.5 and 1), that combines the tetrahedral framework of Ba$_{2}$CoO$_{4}$ with the octahedral coordination geometry of Sr$_{2}$CoO$_{4}$\cite{Matsuno,Wang} with a K$_2$NiF$_4$-type structure. The interest in such an ``hybrid'' phase has originally been motivated by the contrast in the physical properties of the two materials, where Ba$_{2}$CoO$_{4}$ is an antiferromagnetic - insulator,~\cite{Jin} while Sr$_{2}$CoO$_{4}$ is a ferromagnetic - metal.\cite{Matsuno} As will be discussed in the article, this new series crystallizes in an hexagonal structure in which four Co--O polyhedra (three tetrahedra and one octahedron) share their corners to form Co$_4$O$_{15}$ tetrahedral clusters. Several materials with similar structural motifs and chemical formulas Ba$_{6}$\emph{Ln}$_{2}$\emph{M}$_{4}$O$_{15}$ have been first reported over two decades ago by M\"{u}ller-Buschbaum \emph{et al.}.~\cite{Buschbaum} Some of the iron-based congeners have recently been revisited by Abe \emph{et al.}~\cite{Abe} with a renewed focus on their physical properties. In those compounds, Fe$^{3+}$ ions (3$d^5$, S = 5/2) arranged in Fe$_4$O$_{15}$ clusters display a behavior indicative of magnetic tetramers with a ferrimagnetic ground state of the total spin S$_T$ = 5.~\cite{Abe} In our newly synthesized compound, (BaSr)$_{4}$Co$_{4}$O$_{15}$, the Co ions are expected to exhibit a mixed valence of average value 3.5+, consisting of trivalent ($3d^6$) and tetravalent ($3d^5$) states. Given the Co cations distribution on the two non-equivalent sites with the multiplicity ratio 3:1, it appeared as a natural choice to attempt to tune the system properties and induce a charge ordering by a controlled substitution of Sr$^{2+}$/Ba$^{2+}$ with La$^{3+}$. On this basis, one would expect that the charge distribution in (BaSr)$_{3.5}$LaCo$_{4}$O$_{15}$ (where average valence of the Co becomes 3.25+) will favor a $3d^5$ (Co$^{4+}$) state at the octahedral site and  $3d^6$ (Co$^{3+}$) at the tetrahedral site. Furthermore, for the case of (BaSr)$_{3}$La$_{2}$Co$_{4}$O$_{15}$, all Co ions are anticipated to be trivalent (Co$^{3+}$ $3d^6$). Our study reveals that all compositions are characterized by competing ferro- and antiferro-magnetic exchange interactions that adds to the geometrical frustration inherent in the Co sublattice to provide essential ingredients for spin-glass behavior. Instead of charge ordering, we find evidence for disordered charge distribution and an admixture of ligand-hole configurations. Elastic and inelastic neutron scattering measurements reveal that a spin-glass phase transition takes place at approximately 13~K, and evidence a systematic slowing down of spin dynamics upon increasing the La concentration.

\section{Experimental details}

Polycrystalline samples of (BaSr)$_{4-x}$La$_{2x}$Co$_{4}$O$_{15}$ (0 $\leq$ x $\leq$ 1) used in this study were prepared by solid-state reaction. Stoichiometric mixture of BaCO$_3$ (Alfa Aesar 99.99\%),  SrCO$_3$ (Alfa Aesar 99.994\%), Co$_3$O$_4$ (Alfa Aesar 99.99\%) and La$_2$O$_3$ (Alfa Aesar 99.999\%) were heated up to 1100$^\circ$C for 24 hours under ambient pressure. After quenching down to room temperature, the powder was re-ground and repeated with the above process. Single crystal specimens of (BaSr)$_{4}$Co$_{4}$O$_{15}$ were grown by the floating-zone technique using a Cannon MDH20020 image furnace. For all synthesized samples (x = 0, 0.5, and 1) the nominal ratio Sr:Ba was set to be 1:1.
Characterizations conducted by powder X-ray diffraction indicated that the $x$ = 0 and 0.5 samples were free of impurities, and the $x$ = 1 sample contained a small amount of the layered perovskite phase (Sr,Ba,La)$_{2}$CoO$_{4}$. Attempts of synthesizing samples with larger La content, $x >$ 1, were unsuccessful, leading to increased amounts of impurities.

The powder X-ray diffraction patterns were obtained using a Rigaku MiniFlex II powder diffractometer equipped with a Cu target and a Ni-filter for removal of Cu-K$_\beta$ radiation. Single-crystal x-ray diffraction measurements on the x = 0 sample were carried out using a Bruker SMART APEX CCD diffractometer with fine focus Mo K$_\alpha$ radiation ($\lambda$=0.7107~{\AA}). Determination of integrated intensities and a global cell refinement were performed with the Bruker SAINT software package using a narrow-frame integration algorithm. A total number of 5319 reflections was used for the structural analysis. The crystal structure refinement was carried out using SHELXTL software.~\cite{shelxtl}

Neutron powder diffraction measurements were performed at the HB2A high-resolution powder diffractometer, at the High Flux Isotope Reactor (HFIR), using the wavelengths $\lambda$ = 1.539~{\AA} and $\lambda$ = 2.41~{\AA}. This instrument is equipped with 44 $^3$He detector tubes that were scanned to cover the total 2$\Theta$ range of 4$^\circ$-150$^\circ$, in steps of 0.05$^\circ$. More details about the HB2A instrument and data collection strategies can be found in Ref.~\onlinecite{HB2A}. Measurements were made on approximately 7 grams of sample held in a vanadium holder with 0.8~cm diameter. Rietveld refinements were performed using the FULLPROF program.~\cite{fullprof}

The magnetization M(H,T) measurements were carried out using a commercial (Quantum Design) superconducting quantum interference device (SQUID) magnetometer. Zero field cooling (ZFC) and field cooling (FC) magnetization measurements were performed in a magnetic field of 0.1~Tesla in the temperature range 1.8~K $\leq$ T $\leq$ 350~K.

Inelastic neutron scattering experiments reported here were undertaken on approximately 15~g of powder samples, sealed in aluminium holders. The inelastic neutron spectra were collected using the time-of-flight spectrometer BASIS at the SNS.~\cite{BASIS} This spectrometer uses backscattering neutron reflections from Si(111) analyzer crystals to select the final energy of the neutron of 2.08 meV, and allows achieving an energy resolution at the elastic position of about 3.5~$\mu$eV. More details about the BASIS spectrometer are given in Ref.~\onlinecite{BASIS}.

\begin{table*}[tbp]
\caption{\label{table1} Refined structural parameters of (BaSr)$_{4}$Co$_{4}$O$_{15}$ from single-crystal x-ray data collected at 173~K.}
\begin{ruledtabular}
\begin{tabular}{ccccccccccc}
\centering{Atom (Wyck.)} & $x$ & $y$ & $z$ & U11 & U22 & U33 & U23 & U13 & U12 & Occ.(\%)\\[3pt]
\hline
\centering {Sr1/Ba1 ($2a$)}& 0 & 0 & 0.0101(1) & 0.0139(3) & 0.0139(3) & 0.0131(3) & 0.0000 & 0.0000 & 0.0069(1) & 0.91/0.09(1)\\[3pt]
\centering {Sr2/Ba2 ($2b$)}& 1/3 & 2/3 & 0.4993(1) & 0.0178(1) & 0.0178(1) & 0.0124(2) & 0.0000 & 0.0000 & 0.0089(1) & 0.02/0.98(1)\\[3pt]
\centering {Sr3/Ba3 ($6c$)}& 0.4773(1) & 0.5226(1) & 0.8376(1) & 0.0095(1) & 0.0095(1) & 0.0129(2) & -0.000(1) & 0.000(1) & 0.0054(1) & 1.00/0.00(1)\\[3pt]
\centering {Sr4/Ba4 ($6c$)}& 0.1734(1) & 0.8265(1) & 0.1800(5) & 0.0133(1) & 0.0133(1) & 0.0188(1) & -0.0003(1) & 0.0003(1)& 0.0068(1) & 0.02/0.98(1)\\[3pt]
\centering {Co1 ($2b$)}& 1/3 & 2/3 & 0.0268(2) & 0.0090(3) & 0.0090(3) & 0.0137(7) & 0.00000 & 0.00000 & 0.0045(1) & 1.00 \\[3pt]
\centering {Co2 ($6c$)}& 0.1761(1) & 0.8238(1) & 0.6782(1) & 0.0098(2) & 0.0098(2) & 0.0127(3) & 0.0001(1) & -0.0001(1) & 0.0042(2) & 1.00\\[3pt]
\centering {O1 ($12d$)}& 0.6783(2) & 0.0660(2) & 0.0391(4) & 0.009(1) & 0.016(1) & 0.014(1) & -0.005(1) & -0.004(1) & 0.003(1) & 1.00\\[3pt]
\centering {O2 ($6c$)}& 0.2499(1) & 0.7500(1) & 0.8518(6) & 0.012(1) & 0.012(1) & 0.012(1) & 0.0026(7) & -0.0026(7) & 0.004(1) & 1.00\\[3pt]
\centering {O3 ($6c$)}& 0.4102(1) & 0.5897(1) & 0.1745(7) & 0.014(1) & 0.014(1) & 0.016(1) & 0.0005(7) & -0.0005(7) & 0.010(1) & 1.00\\[3pt]
\centering {O4 ($6c$)}& 0.9070(1) & 0.0929(1) & 0.2737(4) & 0.018(1) & 0.018(1) & 0.022(2) & -0.0025(7) & 0.0025(7) & 0.012(1) & 1.00\\
\hline \\
\multicolumn{11}{c}{$P6_3mc$, $a$ = $b$ = 11.6450~\AA, $c$ = 6.8600~\AA, $V_{cell}$ = 805.63~\AA$^{3}$;    R$_{1}$ = 0.0315, wR$_{2}$ = 0.0746, GooF = 1.150}\\
\end{tabular}
\end{ruledtabular}
\end{table*}

\begin{table}[tbp!]
\caption{\label{table2}Selected bond distances d(\AA), and estimated Co valences using BVS in (BaSr)$_{4-x}$La$_{2x}$Co$_{4}$O$_{15}$, at the room temperature}
\begin{ruledtabular}
\begin{tabular}{ccccc}
 & & x = 0 & x = 0.5 & x = 1 \\[3pt]
Co1 &- O3 (x 3) & 1.85(1) & 1.87(1) & 1.89(1)  \\[3pt]
&- O2 (x 3) & 2.08(1) & 2.05(1) & 2.06(1) \\[3pt]
Co2 &- O1 (x 2) & 1.79(1) & 1.81(1) & 1.84(1) \\[3pt]
&- O2 (x 1) & 1.91(1) & 1.95(1) & 1.95(1) \\[3pt]
&- O4 (x 1) & 1.83(1) & 1.77(1)& 1.74(1) \\[3pt]
\hline\\
\multicolumn{2}{l}{d(Co1 - Co2)~ ($J_1$, via O2)} & 3.99(1) & 4.00(1)& 4.00(2)\\[3pt]
\multicolumn{2}{l}{Co1 - O2 - Co2} & 177.1$^{\circ}$ & 178.2$^{\circ}$ & 178.8$^{\circ}$ \\[3pt]
\multicolumn{2}{l}{d(Co2 - Co2)~ ($J_2$, via O1, O4)} & 4.97(1)& 4.91(1)& 4.92(1) \\[3pt]
\multicolumn{2}{l}{d(Co2 - Co2)~ ($J_3$, via O1)} & 5.48(1)& 5.57(1)& 5.57(1) \\[3pt]
\hline\\
BVS & Co1 & 2.92 & 2.87 & 2.75 \\[3pt]
BVS & Co2 & 2.65 & 2.65 & 2.60 \\[3pt]
\end{tabular}
\end{ruledtabular}
\end{table}

\begin{figure*}[tbp]
%\begin{figure}[tbp]
\includegraphics[width=6.4in]{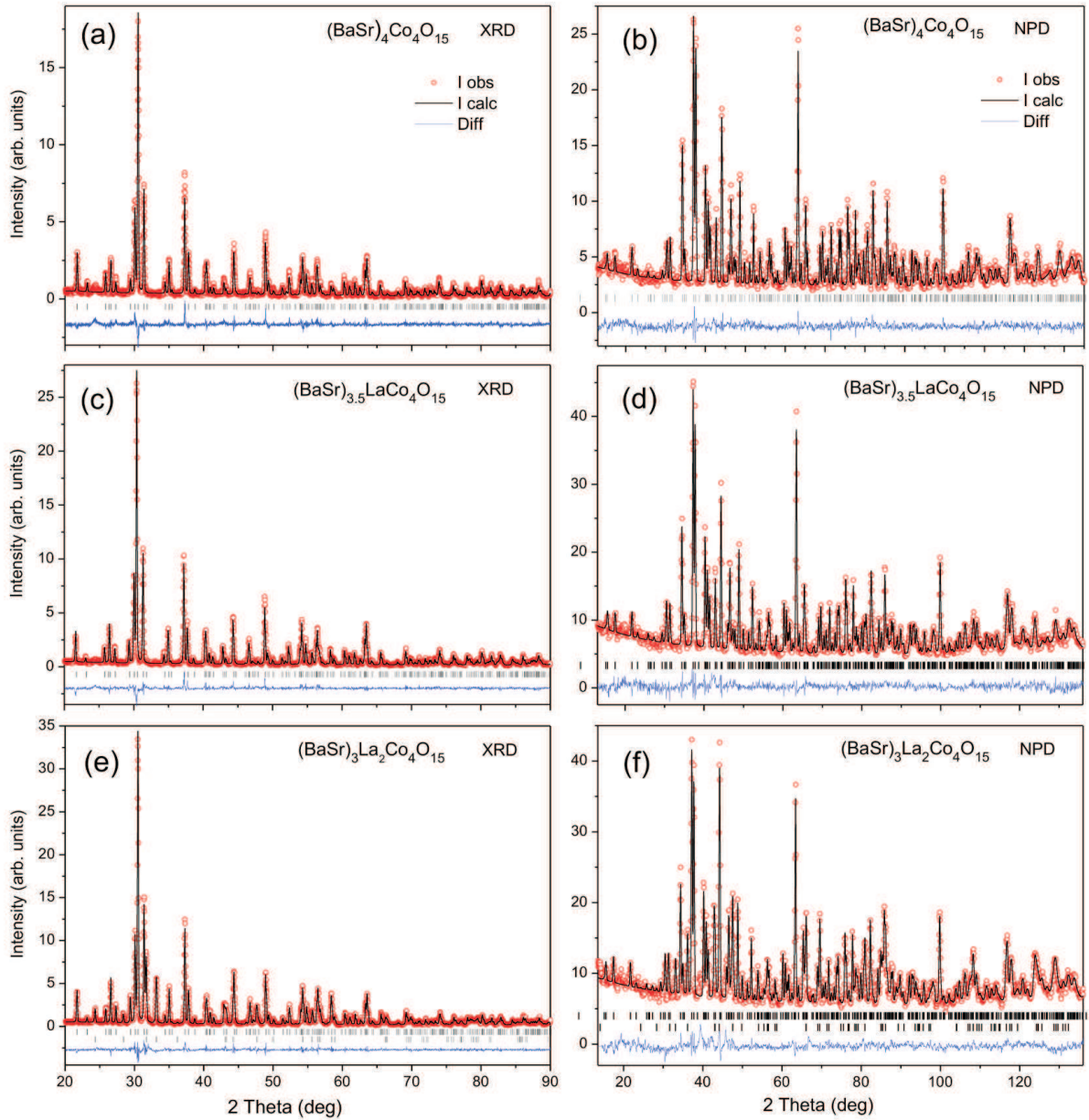}
\caption{\label{diff}(Color online) X-ray and neutron powder diffraction patterns and Rietveld fits of (BaSr)$_{4-x}$La$_{2x}$Co$_{4}$O$_{15}$ ($x$ = 0, 0.5 and 1) at 300 K. The displayed data was collected using a Cu K$_\alpha$ x-ray source (a)(c)(e) and, respectively, using a neutron beam with the wavelength of 1.539~{\AA} (b)(d)(f). The difference between observed and calculated profiles is shown at the bottom of each panel.}
\end{figure*}

\section{Results and Discussion}
\subsection{Crystal structure refinement}

The x-ray investigation of the (BaSr)$_{4}$Co$_{4}$O$_{15}$ single crystal revealed that this compound has an hexagonal structure described by the non-centrosymmetric space group $P6_3mc$. The four Co ions of the unit cell occupy two inequivalent crystallographic sites: $2b$ ($1/3,2/3,z$) and $6c$ ($x,-x,z$). The Co1 ions positioned at the 2$b$ site are coordinated octahedrally by six oxygen atoms (3x O3, 3x O2), while the Co2 ions filling the 6$c$ site are coordinated tetrahedrally by four oxygens (2x O1, O2 and O4). The polyhedra are sharing their corners to form Co$_4$O$_{15}$ clusters with a tetrahedral packing, as displayed in Fig.~\ref{struct}(a). The shortest Co bonds inside the tetrahedron are 3.99(1)~{\AA} linking the Co1 and Co2, while the angle Co1-O2-Co2 is 176.8(1)$^\circ$. The other edges of the tetrahedron, defined by Co2 atoms, measure 5.48(1)~{\AA}. The Co2 atoms located in the vertices of different tetrahedral clusters are linked through double oxygen bridges with a Co2 -- Co2 distance of 4.938(1)~{\AA}. Altogether, the Co atoms form interconnected honeycomb-like layers that are stacked along the $c$-axis direction. A projection of the structure along the $c$-axis is shown in Fig.~\ref{struct}(b). The most relevant magnetic interaction paths are depicted by arrows and labeled as $J1$, $J2$ and $J3$ in Fig.~\ref{struct}(b). $J1$ refers to the exchange interaction between Co1 and Co2 mediated by O2 (intra-cluster interaction), and $J2$ stands for the coupling between Co2 atoms of adjacent clusters (inter-cluster interaction), defining the honeycomb matrix. The third interaction integral $J3$, between Co2 atoms belonging to the same cluster, is expected to be much smaller because of much larger interatomic distance.

The results of our structural analysis indicate that Sr and Ba atoms are distributed over four different sites in an almost ordered manner. The Sr occupies the special site $2a$ ($0,0,z$)(Sr1) and $6c$ (Sr3), whereas the Ba resides at the 2$b$ ($1/3,2/3,z$) site (Ba2) and at the non-equivalent 6$c$ (Ba4). Structural parameters of the (BaSr)$_{4}$Co$_{4}$O$_{15}$ compound and the most important bond distances are shown in Table~\ref{table1} and ~\ref{table2}.

Rietveld refinements of all (BaSr)$_{4-x}$La$_{2x}$Co$_{4}$O$_{15}$ powder samples were carried out using combined powder X-ray and neutron diffraction data sets, in order to extend the number of observations. In addition, neutron diffraction provides an excellent complement to the X-ray due to their greater sensitivity in detecting the oxygen positions. The structural model described previously, based on space group $P6_3mc$, gave a good fit to all experimental data. As mentioned above, the higher doping sample was found to contain an impurity phase, (Sr,Ba,La)$_{2}$CoO$_{4}$, which was included in the refinement. The observed, calculated and difference profiles along with the positions of all contributing $hkl$ reflections are displayed in Figure~\ref{diff}.

Refinements of the oxygen occupancies of the four inequivalent oxygen sites suggest a stoichiometry that is very close to the nominal value O$_{15}$. In the doped samples, the lanthanum atoms were tentatively placed to substitute at any of the Ba/Sr sites and their occupancies were collectively refined. We found, however, that La$^{3+}$ prefers to occupy the Sr3/Ba3 ($6c$) site, which is fully consistent with the structural model proposed for the Fe-analogs Ba$_{6}$\emph{Ln}$_{2}$Fe$_{4}$O$_{15}$.~\cite{Abe} The refined La stoichiometry is 1.4$\pm$0.2 for the nominal $x$ = 0.5 sample, and 2.2$\pm$0.2 for the $x$ = 1 sample.~\cite{Supplemental} On the other hand, the analysis indicates that upon La doping ($x$ = 1), both the $a$- and $c$- axes contract by approximately 0.15 and 0.4\%, respectively. In terms of interatomic distances this translates to an elongation of the Co2--O1 bond whereas the Co2--O4 bond contracts by approximately the same amount. Overall, this produces a distortion of the tetrahedral environment of the Co2, that leads to a shortening of the distance between Co2 atoms from adjacent cluster units (see $J2$ in Fig.~\ref{struct}(b)). The O2 atom shifts slightly towards the Co1 without affecting the intra-cluster distance Co1--Co2.

Bond-valence sum (BVS) calculations were performed according to Brown and Altermatt,~\cite{bvs} using the program VaList.~\cite{valist} The resulting valence values for the two cobalt sites are significantly lower than those expected based on stoichiometry. For the octahedrally coordinated Co1 site the calculated value is close to 3+ (2.92+) for the undoped sample ($x$ = 0), and it progressively decreases to 2.75+ for the $x$ = 1 doping. In contrast, the calculated charge at the tetrahedral Co2 site is 2.6+, and it does not seem to change much with the La doping. Overall, our calculation suggests the possibility of a partial charge transfer from cobalt to the neighboring oxygens. As a result of hybridization between Co $e_g$ and O $2p$ states, the new electronic ground state can be expressed by a linear combination of $d^n$, $d^{n+1}\underline{\mathrm{L}}$ and $d^{n+2}\underline{\mathrm{L}}^{2}$ states, where ``$\underline{\mathrm{L}}$'' stands for a hole at the O $2p$ level. The values of $n$ are 6 and 5 for the Co$^{3+}$ and Co$^{4+}$, respectively. Similar behavior has been reported for other cobaltates, such as LiCoO$_2$,~\cite{Elp} SrCoO$_3$,~\cite{Potze} CaBaCo$_4$O$_7$~\cite{Caignaert}. Certainly, a study by x-ray absorption spectroscopy would be desirable to confirm this hypothesis. For our material, seeing the significantly larger distance between both non-equivalent Co sites and O2 site (see Table~\ref{table2}), one may infer that the hole ($\underline{\mathrm{L}}$) appears prominently on the O2 atom shared by the two cobalt atoms.

\subsection{Macroscopic magnetic measurements}

\begin{figure}[tbp]
\includegraphics[width=3.3in]{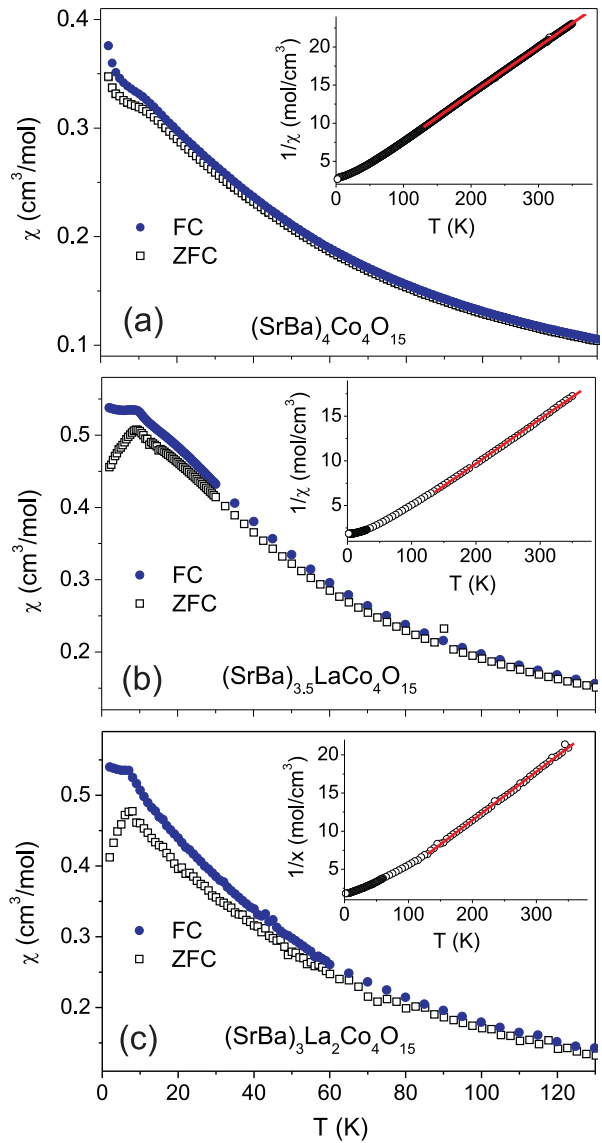}
\caption{\label{susc}(Color online) ZFC and FC susceptibility data for three different doping levels. Insert: Temperature dependence of the inverse susceptibility (1/$\chi$) described by a Curie-Weiss law above approximately 130~K.}
\end{figure}

The temperature dependence of low field (H=0.1~Tesla) static susceptibility ($\chi=M/H$) of (BaSr)$_{4-x}$La$_{2x}$Co$_{4}$O$_{15}$ polycrystalline samples, measured under zero-field cooled (ZFC) and field cooled (FC) conditions, are presented in Fig.~\ref{susc}. For all three specimens ($x$ = 0, 0.5, 1) the susceptibility shows a kink at approximately 10~K, and obeys a Curie-Weiss law $\chi$(T) = C/(T - $\Theta$) at higher temperatures (see solid lines in the insets of Fig.~\ref{susc}). As is readily visible in Fig.~\ref{susc}, the ZFC and FC curves exhibit a clear irreversibility suggesting that a spin-glass phase transition takes place. The bifurcation between the curves becomes more pronounced with increasing La concentration. Furthermore, the splitting point shifts towards higher temperatures indicating that short-range correlations start to develop well above the glassy ordering temperature T$_{g}$.

The temperature dependence of the inverse susceptibility (1/$\chi$), shown in the insert of Fig.~\ref{susc}, was fit over the temperature range 130 - 350 K with a Curie–Weiss formula. The magnetic parameters estimated from $\chi$ versus T plots are summarized in Table~\ref{table3}. Interestingly, the resulting Curie-Weiss temperatures indicate a progressive crossover from dominant antiferromagnetic interactions ($\Theta \approx -20.9~K$) in $x$ = 0 to ferromagnetic interactions ($\Theta \approx 21.3~K$) in $x$ = 1. For intermediate doping level, $x$ = 0.5, the couplings are found to be predominantly ferromagnetic with the Weiss temperature 5.46~K. It is also noteworthy that the Curie-Weiss temperatures are roughly the same as the freezing temperature T$_{g}$, suggesting that the magnetic frustration is not significantly strong in spite of the geometrically frustrated tetrahedral configuration of the nearest neighbor ions. One can then deduce that the spin-freezing phenomena in these materials are mostly due to the spin disorder and to the simultaneous presence of ferro- and antiferromagnetic coupling between the magnetic sites.

The effective magnetic moment ($\mu_{\mathrm{eff}}$) of Co is found to decrease as La concentration increases. It reaches $\simeq$4.7~$\mu_{B}$ for $x$ = 0, and decreases to $\simeq$3.9~$\mu_{B}$ for $x$ = 1. In a tetrahedral environment, Co$^{4+}$ ($3d^5$) and Co$^{3+}$ ($3d^6$) ions are generally expected to exhibit the high spin states S = 5/2 and S = 2, respectively. This is because the tetrahedral sites have, compared to the octahedral sites, an inverted and smaller crystal-field splitting (i.e. the two $e_g$ orbitals are lower in energy than the three $t_{2g}$ orbitals). In contrast, in octahedral fields, both Co$^{4+}$ and Co$^{3+}$ can exhibit either high spin, intermediate (S = 3/2, Co$^{4+}$ and S = 1, Co$^{3+}$) or low spin (S = 1/2, Co$^{4+}$ and S = 0, Co$^{3+}$).~\cite{Candela,Potze} From the estimated magnitude of the effective moment, one can depict a scenario where the tetrahedral sites may be primarily occupied by a high-spin Co$^{3+}$ ions and the octahedral site by intermediate-spin of Co$^{4+}$ and/or Co$^{3+}$ ions. Potze \emph{et al.}~\cite{Potze} have demonstrated using atomic multiplet calculations that the intermediate-spin state can become the ground state due to the relative stability of a ligand-hole configurations, such as $d^{6}\underline{\mathrm{L}}$, for Co$^{4+}$, and $d^{7}\underline{\mathrm{L}}$, for Co$^{3+}$. There is also always the possibility of hybridization between various spin states, and the admixture will depend on La doping.

The charge-transfer model can also be employed to explain the presence of ferromagnetic interactions in the system, with itinerant oxygen holes coupling antiferromagnetically to the $d$ electrons of $e_g$ symmetry of the adjacent Co ions.~\cite{Potze} On the other hand the superexchange mechanism via oxygen, without a hole, could lead to an antiferromagnetic coupling between the nearest neighbor cobalt sites. The coexistence of the two mechanisms resulted from the charge disorder can account for the spin-glass behavior.

\begin{table}[btp!]
\caption{\label{table3} Magnetic ordering temperatures, Curie-Weiss temperatures, and the effective magnetic moments determined from susceptibility measurements.}
\begin{ruledtabular}
\begin{tabular}{cccc}
Sample & T$_{g}$~(K) & $\Theta_{CW}$~(K) & $\mu_{\mathrm{eff}}$ ($\mu_{B}$) \\
\hline\\[1pt]
(BaSr)$_{4}$Co$_{4}$O$_{15}$ & 10(1) & -20.9(2) & 4.8(1) \\[3pt]
(BaSr)$_{3.5}$LaCo$_{4}$O$_{15}$ & 10(1) & 5.2(9) & 4.5(1) \\[3pt]
(BaSr)$_{3}$La$_{2}$Co$_{4}$O$_{15}$ & 10(1) & 21.3(15) & 3.9(1) \\[3pt]
\end{tabular}
\end{ruledtabular}
\end{table}

\subsection{Elastic and inelastic neutron scattering}
\begin{figure}[tbp]
\includegraphics[width=3.5in]{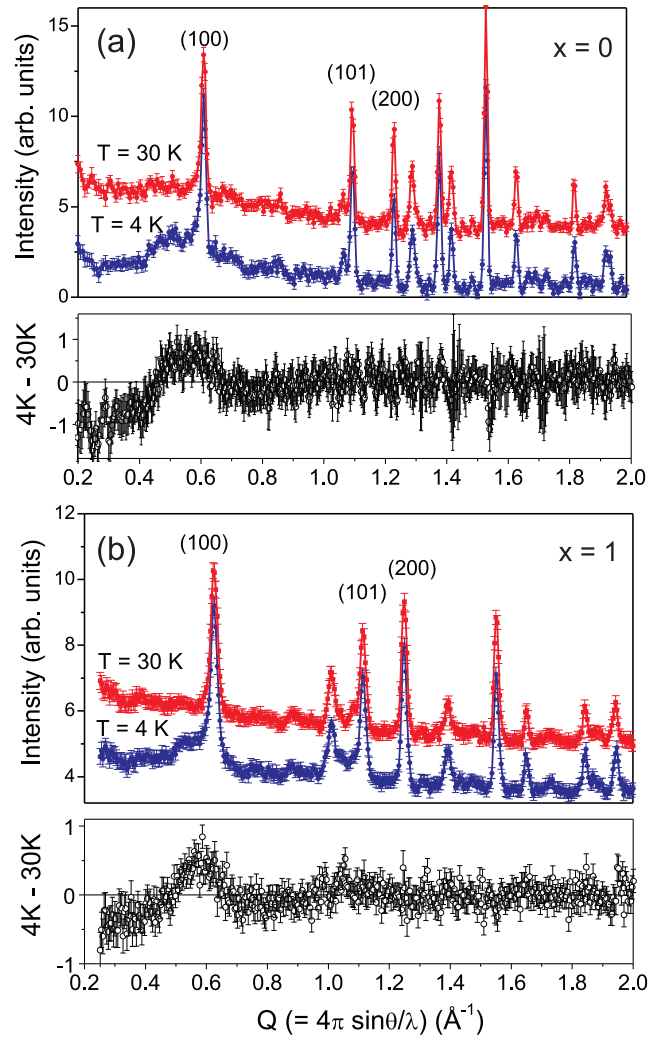}
\caption{\label{npddiff}(Color online) Neutron powder diffraction data collected on (BaSr)$_{4-x}$La$_{2x}$Co$_{4}$O$_{15}$ where $x$ = 0 (a) and 1 (b) at 30~K and 4~K, using 2.41{\AA} wavelength. The 30~K data has been vertically offset for clarity. Lower panels in (a) and (b) show the difference scattering obtained by subtracting the 30 K from data from 4 K data.}
\end{figure}

Neutron powder diffraction measurements have been carried out above and below the transition temperature, T$_{g}$, to monitor the changes in the scattering patterns. The data, shown in Fig.~\ref{npddiff}, provide no evidence for structural changes or magnetic long-range ordering. The difference scattering, resulted from subtracting the 30~K data from the 4.2~K data reveals a distinct broad feature around the (100) peak position. The negative differential signal at $Q <$0.4~\AA$^{-1}$ is caused by the paramagnetic scattering that exists above the ordering temperature. At higher $Q$ ($>$1.4~\AA$^{-1}$), the differential scattering is dominantly produced by the small shift of peak positions due to the lattice compression. This observation is consistent with the spin-glass-like behavior, suggested on the basis of the magnetic susceptibility. It is also worth noting that although the two specimens ($x$ = 0 and $x$ = 1) exhibit different dominant magnetic interactions in the paramagnetic state, the $Q$ dependence of the magnetic diffuse scattering from their spin-glass ground state looks remarkably similar.

\begin{figure}[tbp]
\includegraphics[width=3.4in]{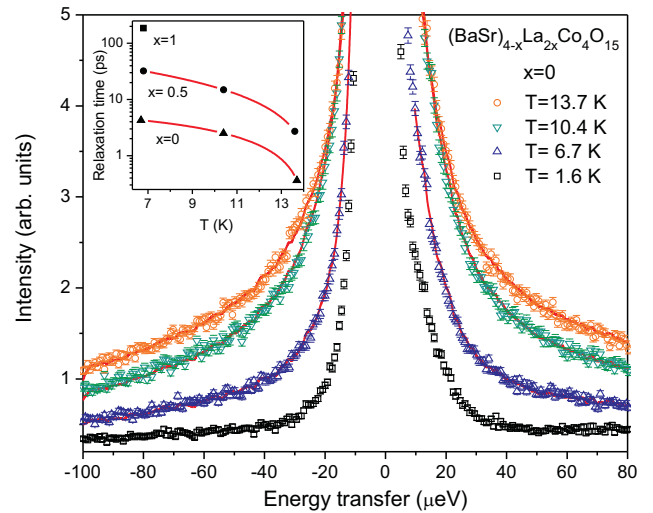}
\caption{\label{INSPlotCut}(Color online) High-resolution inelastic neutron scattering data of (BaSr)$_{4}$Co$_{4}$O$_{15}$, collected at various temperatures across the spin-glass transition. The solid line is a fit to the data, as described in the text. Inset: Variation of relaxation time with temperature in the undoped and La doped samples. The solid line is drawn as a guide to the eye.}
\end{figure}

To get deeper insight into spin dynamics and relaxation phenomena across the glassy transition region, we performed high-resolution inelastic neutron scattering (INS) measurements using the backscattering spectrometer BASIS at SNS. Figure~\ref{INSPlotCut} shows intensities (integrated over the range $Q \approx$ 0.25 to 2 \AA$^{-1}$) versus neutron energy transfer, at various temperatures. The selected $Q$ range is compatible with the relaxed $Q$ resolution of the spectrometer and it provides suitable data for increasing the statistics of the scattering signal.~\cite{Supplemental} The quasi-elastic peak at T = 1.6~K has a full width at half maximum (FWHM) of 3.3(2) $\mu$eV, consistent with the instrumental energy resolution. There is a clear narrowing of the line shape as the sample temperature is lowered below the transition point, which is a signature of a dynamical magnetic transition to a glass-like state. The low-frequency inelastic intensity weakens upon cooling, and at the same time, the elastic intensity increases indicating the freezing of spin correlations over finite length scale. To evaluate quantitatively the temperature dependence of the spin relaxation rate, the quasielastic scattering was modeled using the Fourier transform of a stretched exponential relaxation function of the form: $S(Q,t) = \exp[-(t/\tau)]^{\beta}$, where $\beta$ represents the stretching exponent $(0<\beta\leq{1})$, and $\tau$ the relaxation time. The stretched exponential function has traditionally been used to model relaxation in spin glasses or structural glasses, and a generalization has been discussed recently.\cite{Palmer,Pickup} In the modeling procedure, the Fourier Transform (time to energy) of the stretched exponential relaxation function was computed with sufficient precision using a numerical procedure adapted to the particular shape of the function.\cite{Vooren} The $\beta$ parameter was set to 0.25 and held fixed throughout, because, as it is a phenomenological parameter, a possible temperature dependence will add too much uncertainty to the fitting to give dependable results for the relaxation time.  The Fourier Transform was numerically convoluted with the dataset collected at 1.6 K which was used as the elastic reference scattering. The mean spin relaxation time determined at three different temperatures is displayed in the inset of Figure~\ref{INSPlotCut}. For the $x$ = 1 sample, INS data was collected with sufficient statistics at only one temperature, 6.8 K, and therefore only one point is shown. The spin relaxation time exhibits a smooth trend gradually increasing with decreasing the temperature by approximately a factor of 12. Modeling of the data for all three samples reveals a clear tendency towards slower dynamics when La doping increases. For instance, the $\tau$ for the $x$ = 0 sample at 6.8~K was estimated to be 4~ps, while for the $x$ = 1 sample is 187~ps. This is an intrinsic feature, since the freezing temperature has been determined to be about the same in all samples, and is likely an effect of the reduction in charge randomness in the doped samples that leads to stronger correlations between magnetic spins.

\begin{figure}[tbp]
\includegraphics[width=3.4in]{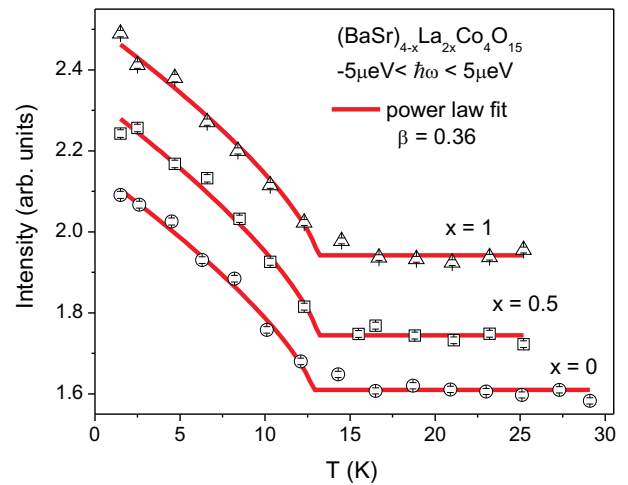}
\caption{\label{orderparam} Temperature dependence of the elastic scattering, from all three samples, integrated over the $Q$ range 0.25-2~\AA$^{-1}$. The solid line represents a power-law fit to the data.}
\end{figure}

Figure~\ref{orderparam} displays the temperature dependence of the elastic scattering, considered as the momentum transfer averaged ($0.25 < Q \leq 2 \AA^{-1}$) integrated intensity over an energy transfer range -5$\mu$eV$\leq\hbar\omega\leq$5$\mu$eV. The scattering amplitude of all three samples follow similar trends and the glassy transition seems to appear at approximately same temperature, in agreement with the susceptibility measurements. To assess the character of the magnetic transition, we determined the critical exponent $\beta$ associated with the magnetic order parameter defined as $I\propto(T_{g}-T)^{2\beta}$. Least-square fits to the data near the transition yielded an exponent $\beta\simeq$~0.36(1) for the undoped system. The estimated values of the critical exponent $\beta$ are very close to those of the 3D Heisenberg universality class, and are consistent with three-dimensional topology of the magnetic interactions within the lattice.

\section{Summary}

To summarize, we have studied the structure and magnetic properties of a new class of cobaltates (BaSr)$_{4-x}$La$_{2x}$Co$_{4}$O$_{15}$, that contains cobalt ions in both tetrahedral and octahedral coordination environments. The existence of two distinct crystallographic sites with a significantly different coordination by oxygen anions, leads to the possibility of coexistence of multiple valence states and spin configurations. Within the crystal structure, three Co--O tetrahedra and one octahedron are linked together by shared oxygens to form Co$_4$O$_{15}$ tetrahedra clusters. The magnetic exchange interaction pattern can be defined in terms of two main exchange paths (one intra-cluster ($J1$) and one inter-cluster ($J2$)) and establishes an interconnected honeycomb-like network. Spin-glass behavior is evidenced in all studied samples by the irreversibility between FC and ZFC magnetization and the absence of magnetic reflections in the neutron diffraction pattern at low temperatures. Detailed structural analysis followed by bond-valence-sum calculations, as well as the magnetic parameters suggested the possibility of a partial charge transfer from cobalt to the neighboring oxygens and the presence of ligand-hole states that stabilize an intermediate-spin-state configuration at the octahedral site. The Curie-Weiss behavior at high temperatures, above the spin-freezing transition, reveals a progressive crossover from dominant antiferromagnetic interactions in the undoped sample ($x$ = 0) to ferromagnetic interactions in the doped samples ($x$ = 0.5, 1). However, all samples are subject to a random charge distribution which produces competition between ferro- and antiferro-magnetic correlations, and consequently, spin-glass behavior. Our neutron inelastic scattering study indicated that the transition into the spin-glass state belongs to the universality class of the 3D classical Heisenberg model. In addition, our data showed a clear trend of slowing down of spin-dynamics upon increasing La concentration, pointing to a reduction in charge randomness in the doped samples.

An interesting future direction of this project would involve partial substitution of cobalt by other metal transition ions that would adjust the degree of Co valence mixing and finely tune the magnetic interactions to create other magnetically ordered states.

\begin{acknowledgments}
Authors would like to thank H. Sha and J. Zhang for their interest and support at the early stage of this work. Research at Oak Ridge National Laboratory was sponsored by the Scientific User Facilities Division, Office of Basic Energy Sciences, U. S. Department of Energy. RJ acknowledges the financial support by ORNL Neutron Sciences Visitor Program and DMR-1002622. DBM was supported by LSU the Chancellor's Future Leaders in Research Program.
\end{acknowledgments}


\begin{thebibliography}{}
\bibitem{Yamada} K. Yamada, M. Matsuda, Y. Endoh, B. Keimer, R. J. Birgeneau, S. Onodera, J. Mizusaki, T. Matsuura, and G. Shirane, Phys. Rev. B \textbf{39}, 2336 (1989).
\bibitem{Moritomo} Y. Moritomo, K. Higashi, K. Matsuda, and A. Nakamura, Phys. Rev. B \textbf{55}, R14725 (1997).
\bibitem{Zaliznyak} I. A. Zaliznyak, J. P. Hill, J. M. Tranquada, R. Erwin, and Y. Moritomo, Phys. Rev. Lett. \textbf{85}, 4353 (2000).
\bibitem{Chichev} A. V. Chichev, M. Dlouh\'{a}, S. Vratislav, K. Kn\'{\i}\v{z}ek, J. Hejtm\'{a}nek, M. Mary\v{s}ko, M. Veverka, Z. Jir\'{a}k, N. O. Golosova, D. P. Kozlenko, and B. N. Savenko, Phys. Rev. B \textbf{74}, 134414 (2006).
\bibitem{Senaris} M. A. Se\~{n}aris-Rodriguez and J. B. Goodenough, J. Solid State Chem. \textbf{118}, 323 (1995).
\bibitem{Asai} K. Asai, A. Yoneda, O. Yokokura, J. M. Tranquada, and G. Shirane, J. Phys. Soc. Jpn. \textbf{67}, 290 (1998).
\bibitem{Caciuffo} R. Caciuffo, D. Rinaldi, G. Barucca, J. Mira, J. Rivas, M. A. Se\~{n}aris-Rodriguez, P. G. Radaelli, D. Fiorani, and J. B. Goodenough, Phys. Rev. B \textbf{59}, 1068 (1999).
\bibitem{Phelan} D. Phelan, Despina Louca, S. Rosenkranz, S.-H. Lee, Y. Qiu, P. J. Chupas, R. Osborn, H. Zheng, J. F. Mitchell, J. R. D. Copley, J. L. Sarrao, and Y. Moritomo, Phys. Rev. Lett. \textbf{96}, 027201 (2006).
%\bibitem{Mattausch} Hj. von Mattausch and Hk. Muller-Buschbaum, Z. Anorg. Allg. Chem. \textbf{386}, 1 (1971).
\bibitem{Jin} R. Jin, Hao Sha, P. G. Khalifah, R. E. Sykora, B. C. Sales, D. Mandrus, and Jiandi Zhang, Phys. Rev. B \textbf{73}, 174404 (2006).
\bibitem{Boulahya} K. Boulahya, M. Parras, J. M. González-Calbet, U. Amador, J. L. Martínez, and M. T. Fernández-Díaz, Chem. Mater. \textbf{18} (16), 3898 (2006).
\bibitem{Sofin} M. Sofin, E.-M Peters, M. Jansen Journal of Solid State Chemistry, \textbf{177}, 2550 (2004).
\bibitem{Valldor} M. Valldor and M. Andersson, Solid State Sci. \textbf{4}, 923 (2002).
\bibitem{Maignan} A. Maignan, V. Caignaert, D. Pelloquin, S. H\'{e}bert, V. Pralong, J. Hejtmanek, and D. Khomskii, Phys. Rev. B \textbf{74}, 165110 (2006).
\bibitem{Chapon} L. C. Chapon, P. G. Radaelli, H. Zheng, and J. F. Mitchell, Phys. Rev. B \textbf{74}, 172401 (2006).
\bibitem{Soda} M. Soda, Y. Yasui, T. Moyoshi, M. Sato, N. Igawa, and K. Kakurai, J. Phys. Soc. Jpn. \textbf{75}, 054707 (2006).
\bibitem{Caignaert} V. Caignaert, V. Pralong, V. Hardy, C. Ritter, and B. Raveau, Phys. Rev. B \textbf{81}, 094417 (2010).
\bibitem{Withers} R. L. Withers, M. James, and D. J. Goossens, J. Solid State Chem. \textbf{174}, 198 (2003).
\bibitem{Istomin} S. Y. Istomin, J. Grins, G. Svensson, O. A. Drozhzhin, V. L. Kozhevnikov, E. V. Antipov, and J. P. Attfield, Chem. Mat. \textbf{21}, 4012 (2003).
\bibitem{Khalyavin} D. D. Khalyavin, L. C. Chapon, E. Suard, J. E. Parker, S. P. Thompson, A. A. Yaremchenko, and V. V. Kharton, Phys. Rev. B \textbf{83}, 140403(R) (2011), and refs. therein.
\bibitem{Matsuno} J. Matsuno, Y. Okimoto, Z. Fang, X. Z. Yu, Y. Matsui, N. Nagaosa, M. Kawasaki, and Y. Tokura, Phys. Rev. Lett. \textbf{93}, 167202 (2004).
\bibitem{Wang} X. L. Wang and E. Takayama-Muromachi, Phys. Rev. B \textbf{72}, 064401 (2005).
\bibitem{Buschbaum} H. Mevs, Hk. M\"{u}ller-Buschbaum, J. Less-common Met. \textbf{157} 173(1990) and Z. Anorg. Allg. Chem. \textbf{584}, 114 (1990).
\bibitem{Abe} K. Abe, Y. Doi, Y. Hinatsu, K. Ohoyama, Chem. Mater. \textbf{18}, 785, (2006)and J. Solid State Chem. \textbf{182}, 273(2009).
\bibitem{sir2004} M. C. Burla, R. Caliandro, M. Camalli, B. Carrozzini, G. L. Cascarano, L. De Caro, C. Giacovazzo, G. Polidori, and R. Spagna, J. Appl. Cryst., \textbf{38}, 381 (2005).
\bibitem{shelxtl} SHELXTL 6.12; Bruker AXS, Inc., Madison, WI, (1997).
\bibitem{fullprof} J. Rodriguez-Carvajal, Physica B \textbf{192}, 55 (1993). Program available at www.ill.fr/dif/Soft/fp/.
\bibitem{HB2A} V. O. Garlea, B. C. Chakoumakos, S. A. Moore, G. B. Taylor, T.Chae, R. G. Maples, R. A. Riedel, G. W. Lynn, and D. L. Selby, Appl. Phys. A \textbf{99}, 531 (2010).
\bibitem{BASIS} E. Mamontov and K. W. Herwig, Review of Scientific Instruments, \textbf{82}, 085109 (2011).
\bibitem{Supplemental} See Supplemental Material at [URL will be inserted by publisher] for the complete list of structural parameters resulted from Rietveld analysis, and for the $Q$ dependence of the inelastic scattering.
\bibitem{bvs} I. D. Brown and D. Altermatt, Acta Cryst. B \textbf{41}, 244 (1985).
\bibitem{valist} A. S. Wills, VaList, Program available from www.ccp14.ac.uk.
\bibitem{Elp} J. van Elp, J. L. Wieland, H. Eskes, P. Kuiper, G. A. Sawatzky, F. M. F de Groot, T. S. Turner, Phys. Rev. B \textbf{44}, 6090 (1991).
\bibitem{Candela} G. A. Candela, A. H. Kahn, T. Negas, J. Solid State Chem. \textbf{7}, 360, (1973).
\bibitem{Potze} R. H. Potze, G. A. Sawatzky, and M. Abbate, Phys. Rev. B \textbf{51}, 11501 (1995).
\bibitem{Palmer} R. G. Palmer, D. L. Stein, W. Abrahams, and P. W. Anderson, Phys. Rev. Lett. \textbf{53}, 958 (1984).
\bibitem{Pickup} R. M. Pickup,  R. Cywinski, C. Pappas, B. Farago, and P. Fouquet, Phys. Rev. Lett. \textbf{102}, 097202 (2009) and references therein.
\bibitem{Vooren} A. I. van de Vooren and H. J. van Linde, Mathematics of Computation \textbf{20}, 232 (1966).
\end{thebibliography}
\end{document}